\documentclass[10pt,letterpaper]{article}
\usepackage{opex3}

\usepackage{here}
\newcommand\pictc[5]{\begin{figure}
                       \centerline{
                       \includegraphics[width=#1\columnwidth]{#3}}
                   \protect\caption{\protect\label{fig:#4} #5}
                    \end{figure}            }
\newcommand\pict[4][.7]{\pictc{#1}{!tb}{#2}{#3}{#4}}
\newcommand\rpict[1]{\ref{fig:#1}}
\newcommand\leqt[1]{\protect\label{eq:#1}}
\newcommand\reqtn[1]{\ref{eq:#1}}
\newcommand\reqt[1]{(\reqtn{#1})}
\newcounter{Fig}

\begin{document}

\begin{sloppy}

\title{Excitation of guided waves in layered structures with negative refraction}

\author{Ilya V. Shadrivov$^1$, Richard W. Ziolkowski$^2$, Alexander A. Zharov$^{1,3}$, and Yuri S. Kivshar$^1$}

\address{$^1$ Nonlinear Physics Center and Centre for
Ultra-high bandwidth Devices for Optical Systems (CUDOS), Research
School of Physical Sciences and Engineering, Australian National
University, Canberra ACT 0200, Australia}

\address{$^2$ Department of Electrical and Computer
Engineering, University of Arizona, Tucson, AZ 85721, USA}

\address{$^3$ Institute for Physics of Microstructures,
Russian Academy of Sciences, Nizhny Novgorod 603950, Russia}

\email{ivs124@rsphysse.anu.edu.au}

\homepage{http://wwwrsphysse.anu.edu.au/nonlinear}

\begin{abstract}
We study the electromagnetic beam reflection from layered
structures that include the so-called {\em double-negative
materials}, also called {\em left-handed metamaterials}. We
predict that such structures can demonstrate a giant lateral
Goos-H\"anchen shift of the scattered beam accompanied by
splitting of the reflected and transmitted beams due to the
resonant excitation of surface waves at the interfaces between the
conventional and double-negative materials as well as due to
excitation of leaky modes in the layered structures. The beam
shift can be either positive or negative, depending on the type of
the guided waves excited by the incoming beam. We also perform
finite-difference time-domain simulations and confirm the major
effects predicted analytically.
\end{abstract}


\section{Introduction}

An optical beam totally reflected from an interface between two
transparent dielectric media is known to experience a lateral
displacement from the point of reflection predicted by geometric
optics because each plane wave component of the beam undergoes a
different phase change. Such lateral beam shift is usually called
the Goos-H\"anchen effect~\cite{Goos:1947-333:AP}; it occurs at
the angles close to the angle of the total internal reflection by
the interface, and the beam shift is usually much less than the
beam width.

However, much larger beam shifts are known to occur in the layered
structures that support {\em surface or guided waves} which, when
excited, are able to transfer the incoming beam energy along the
interface. Such guided waves are not excited in the case of a
single interface separating two dielectric media because the phase
matching condition between the incident beam and surface waves
cannot be fulfilled. However, guided waves can be excited in
layered structures when the beam is incident at an angle larger
than the angle of the total internal reflection. In this case, the
guided waves are also termed {\em leaky
waves}~\cite{Tamir:1975-273:NRO}. Excitations of leaky waves by
scattering of electromagnetic waves are usually realized in two
well-known geometries, which are used in the solid-state
spectroscopy, also known as the attenuated (or frustrated) total
internal reflection experiments. These two excitation geometries
are: (i) glass prism-air-dielectric structure, usually called {\em
Otto configuration} \cite{Otto:1968-398:ZP} and (ii) prism-
dielectric film -air structure, usually called {\em Kretchmann
configuration} (see, e.g.
Ref.~\cite{Otto:1975-677:OpticalProperties} and references
therein).

Recent experimental fabrication~\cite{Shelby:2001-77:SCI} and
extensive theoretical studies~\cite{Pendry:2004-37:PT} of the
properties of novel microstructured materials with negative
refraction, called {\em double-negative (DNG) materials}, as well
as {\em left-handed metamaterials}, demonstrate that an interface
between the conventional and DNG media can support {\em surface
waves} (or {\em surface polaritons}) of both TE and TM
polarizations~\cite{Ruppin:2000-61:PLA,Shadrivov:2004-16617:PRE}.
Existence of surface waves for all parameters of the interface
suggests that they can enhance dramatically the value of the
Goos-H\"anchen effect for the beam reflection because such surface
waves can transfer the energy of the incoming beam along the
interface, as was first discussed in
Ref.~\cite{Shadrivov:2003-2713:APL}.

The purpose of this paper is twofold. First, we study analytically
the manifestation of the giant Goos-H\"{a}nchen effect that can be
observed in the beam reflection by two types of layered structures
that include DNG media. In these geometries, we demonstrate
resonant excitation of (i) surface waves at a surface of a DNG
medium, and (ii) leaky waves in a structure with a DNG slab. We
study the corresponding stationary guided modes in the layered
structures, and then demonstrate, by solving the stationary
scattering problem, the resonant reflection and transmission due
to the excitation of leaky waves in two different geometries.
Second, we use the direct numerical finite-difference time-domain
(FDTD)
simulations~\cite{Ziolkowski:2003-662:OE,Ziolkowski:2003-1596:OE}
to study the temporal dynamics of the beam scattering and surface
wave excitation, and confirm the major predictions of our theory.

\section{Lateral beam shift}

First, we recall that, in the framework of the scalar theory of
the linear wave scattering~\cite{Brekhovskikh:1980:Waves}, the
lateral shift $\Delta_{r}$ of the beam reflected by a layered
dielectric structure can be defined as follows,
\begin{equation}\leqt{shift_phase}
\Delta_{r} = \frac{d \Phi_{r}}{dk_x},
\end{equation}
where the index `$r$' refers to the beam reflection and $\Phi_{r}$
is the phase of the reflection coefficient. The approximation
Eq.~\reqt{shift_phase} is obtained with the assumptions that the
beam experiences total internal reflection and that the phase of
the reflection coefficient $\Phi_{r}$ is a linear function of the
wave vector component $k_x$ across the spectral width of the beam.

This lateral beam shift and the Goos-H\"{a}nchen effect have been
calculated for several cases of beam reflection from layered
structures with DNG materials, in particular, for the beam
reflection from a single interface~\cite{Berman:2002-67603:PRE},
\cite{Ziolkowski:2003-662:OE} and a periodic structure of
alternating right- and left-handed
layers~\cite{Shadrivov:2003-3820:APL}. Also, the shift of the beam
transmitted through a DNG slab has been studied theoretically as
well in Ref.~\cite{Chen:2004-66617:PRE}.

However, if the phase $\Phi_{r}$ is not a linear function of the
wave number $k_x$ across the spectral width of the beam (e.g., for
narrow beams with wide spectrum), the approximate
formula~\reqt{shift_phase} for the shift of the beam as whole,
strictly speaking, is not valid. In such a case, one can find the
structure of both reflected and transmitted beams as follows,
\begin{equation}
\leqt{reflection_field} E_{r,t}(x) =
\frac{1}{2\pi}\int_{-\infty}^{\infty} \left\{R(k_x),
T(k_x)\right\} \bar{E}_i(k_x) \, dk_x,
\end{equation}
where $\bar{E}_i$ is the Fourier spectrum of the incident beam,
and then define the relative shift of the beams, $\Delta_{r,t}$,
by using the normalized {\em first moment} of the electric field
of the reflected and transmitted beams, $\Delta_{r,t} =
\Delta_{r,t}^{(1)}$, where
\begin{equation}
\leqt{shift}
    \Delta_{r,t}^{(n)} = \frac{\int_{-\infty}^{\infty} x^n
    |E_{r,t}(x)|^2dx}{ a^n
\int_{-\infty}^{\infty} |E_{r,t}(x)|^2 dx},
\end{equation}
where $a$ is the width of the incident beam.

As a matter of fact, the transverse structure of the reflected and
transmitted beams can have a complicated form; and, in general, it
can be asymmetric so that the shift defined by Eq.~\reqt{shift}
may differ essentially from the value following from
Eq.~\reqt{shift_phase}.

The case $\Delta \ll 1$ corresponds to the beam shift much smaller
then the beam width, whereas the case $\Delta \ge 1$ is much more
interesting, and it corresponds to the so-called giant
Goos-H\"{a}nchen effect. The second moment of the reflected and
transmitted beams, $\Delta_{r,t}^{(2)}$, defined by
Eq.~\reqt{shift}, characterizes a relative width of, respectively,
the reflected and transmitted beams, \[ W_{r,t}=
\sqrt{\Delta_{r,t}^{(2)}}.
\]

In what follows, we assume, without specific restrictions of
generality, that the interface between the first and the second
medium is located at $z=0$ and that the incident beam is Gaussian
and has the beam width $a$, i.e., at the interface the electric
field of the beam has the form $E_i(x,z=0) = \exp{(-x^2/4a^2 -i
k_{x0} x)}$. The angle of incidence, $\phi$, of the beam is
defined with respect to the normal to the interface so that the
wave number component along the interface in the medium from which
the beam is incident is $k_{x0} = k_1 \sin{\phi}$ and the
corresponding wave number in the medium into which the transmitted
beam propagates is $k_1 = \omega \sqrt{(\epsilon_1 \mu_1)}/c$.

\section{Excitation of surface waves}

We consider a two-dimensional, three-layered structure
schematically depicted in Fig.~\rpict{sw_geom}(a), where the input
beam, is incident from an optically dense medium (the first
medium) with $\epsilon_1 \mu_1> \epsilon_2 \mu_2$ at an incident
angle larger than the angle of total internal reflection.  Medium
2 represents a gap layer of width $d$ that separates Medium 1 and
3. We assume that the third medium consists of a DNG metamaterial
which possesses both negative real parts of the dielectric
permittivity $\epsilon_3$ and magnetic permeability $\mu_3$. The
interface between medium one and two generates reflected and
transmitted beams. The interface between medium two and three can
support surface waves which are excited resonantly when the
tangential component of the wave vector of the incident beam
coincides with the propagation constant of the corresponding
surface polariton. In such a case, the surface wave can transfer
the energy along the interface leading to an effective enhancement
of the lateral shift of the reflected and transmitted beams.

\pict{fig01}{sw_geom}{Schematic geometry of the excitation of
surface waves in a three-layer structure that includes a DNG
medium.}

In the geometry shown in Fig.~\rpict{sw_geom}, the reflection
coefficient $R= R(k_x)$ for the TE-polarized monochromatic
[$\sim\exp{(i\omega t)}$] plane wave is defined as
\begin{equation} \leqt{reflection3}
R = \frac{( \alpha_1 + 1 )(\alpha_2 + 1 )
            -(\alpha_1 - 1 )( \alpha_2 - 1 )e^{2ik_{z2}d}}{
            (\alpha_1 - 1 )(\alpha_2 + 1 )
            -(\alpha_1 + 1)(\alpha_2 - 1 )e^{2ik_{z2}d}},
\end{equation}
where $\alpha_{1} = k_{z1} \mu_2 /k_{z2} \mu_{1}$, $\alpha_{2} =
k_{z3} \mu_2 /k_{z2} \mu_{3}$, $k_{zi} = (\omega^2 \epsilon_i
\mu_i /c^2 - k_x^2)^{1/2}$, for $i =1,2,3$, and $c$ is the speed
of light in vacuum. For definiteness, we consider only the case of
TE polarized waves, but our studies indicate that the results are
qualitatively similar for the case of TM polarized waves.

Using Eq. \reqt{reflection3}, we can show that the phase of the
reflection coefficient has an abrupt change when $k_x$ coincides
with the wave vector of a surface wave supported by the interface
between medium two and three. Thus, larger values of the lateral
beam shift are expected at angles of incidence for which the beam
spectrum contains the wave vector components having the same $k_x$
as the propagation constant of the surface waves. As was shown
recently, both {\em forward} and {\em backward} surface polaritons
can exist at the DNG interface~\cite{Shadrivov:2004-16617:PRE},
depending on the effective parameters $X =|\epsilon_3|/\epsilon_2$
and $Y = |\mu_3|/\mu_2$. Excitation of the forward surface waves
results in the energy transfer in the direction of incidence. A
negative shift of the reflected beam will be observed for the
excitation of the backward surface waves, this case corresponds to
the conditions $XY>1$ and $Y<1$.

We chose the following parameters for the media in our
three-layered structure: $\epsilon_1 = 12.8$, $\mu_1 = \epsilon_2
= \mu_2 = 1$, $\epsilon_3 = -3$, $\mu_3 = - 0.5$. The propagation
constant of the surface waves, $h$, is found from the relation
\[
h^2 = \epsilon_2 \mu_2 \frac{\omega^2}{c^2}
\frac{Y(Y-X)}{(Y^2-1)},
\]
and for this case the surface waves at the interface are backward
propagating.  Figures~\rpict{sw_angle_gap}(a,b) show the
dependence of the relative beam shift $\Delta$ and the beam width
$W$ on the angle of incidence when $a/\lambda_0 =100/2\pi$ and
$d/\lambda_0 = 3/2\pi$, where $\lambda_0$ is the free-space
wavelength. A distinctive resonant dependence of the beam shift is
observed, and the maximum of this shift corresponds to the phase
matching condition $k_{x0} = h$.

In the beam profiles shown in Fig.~\rpict{sw_losses}, we observe
that the reflected beam has a distinctive double-peak structure.
The first peak corresponds to a mirror reflection, while the
second peak is shifted relative to the point of incidence.  The
latter can be explained by the excitation of surface waves. At the
resonance, this lateral beam shift becomes larger than the width
of the beam itself. The double-peak structure appears only for
relatively narrow beams for which the beam spectrum is wider than
the spectral width of the surface wave mode, the latter can be
found as the width of the resonance shown in
Fig.~\rpict{sw_angle_gap}(a). The components of the beam spectrum
outside this region are reflected in the usual mirror-like
fashion.  The spectral components of the beam near the resonance
transform into an excited surface wave, and they are responsible
for the appearance of the second peak in the shifted reflected
beam. For wider beams, such that their spectrum completely falls
into the spectral region of the surface wave mode, only the
shifted peak appears. With an increase of the beam width, though,
the relative beam shift decreases due to the fact that the
absolute shift of the beam grows slower than the beam width.

\pict{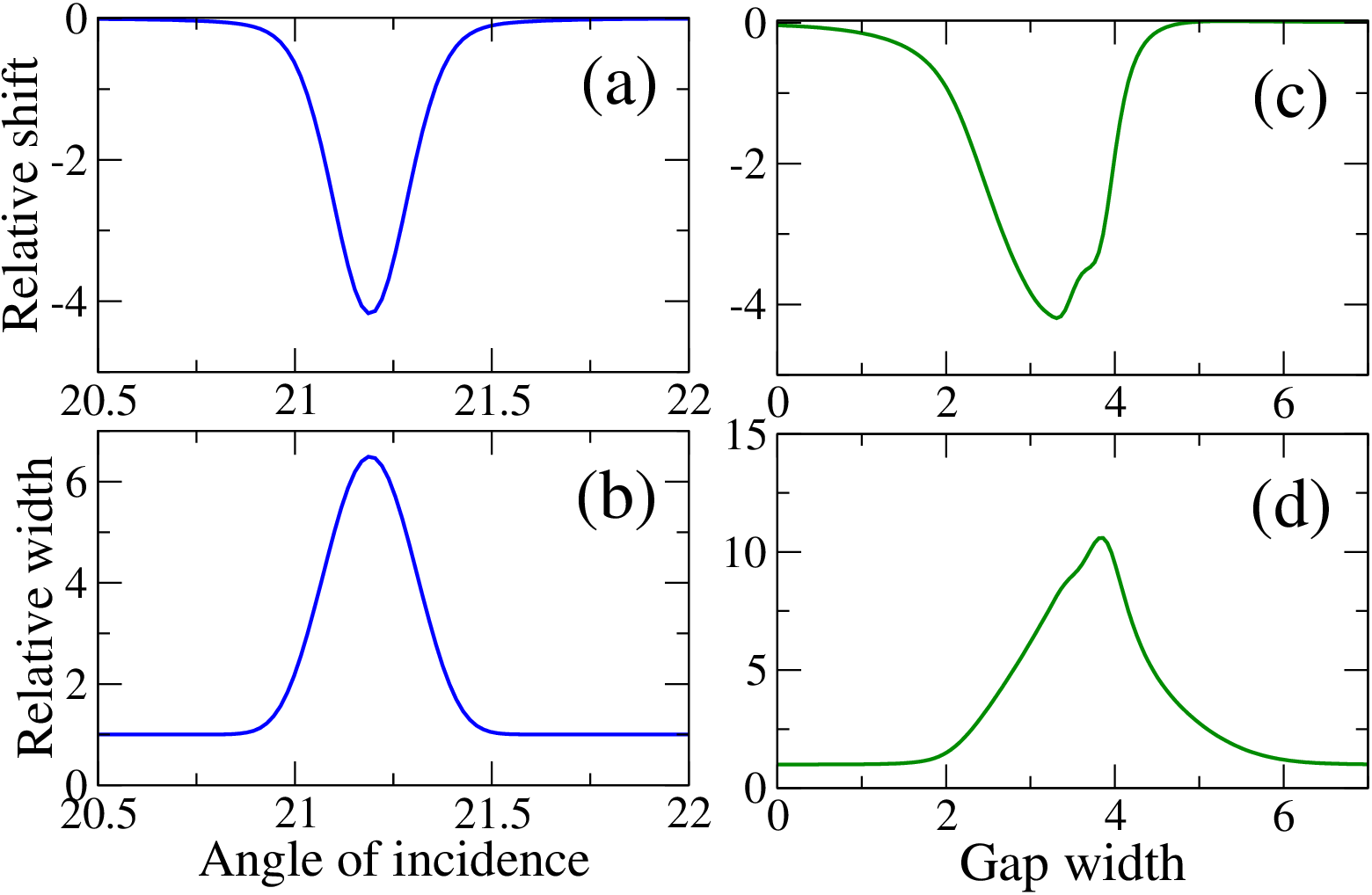}{sw_angle_gap}{ (a,b) Relative beam shift and
beam width vs.  incidence angle. (c,d) Relative shift and width of
the reflected beam vs. normalized gap $2\pi d/\lambda$ at
$a/\lambda = 100/2\pi$. In plots (c,d) the angle of incidence
corresponds to the point of maximum shift in (a).}

\pict{fig03}{sw_losses}{ (a) Relative beam shift and (b)
reflection coefficient vs. the imaginary part of the dielectric
permittivity, parameters are $a/\lambda=100/2.\pi$ and $d/\lambda
= 3/2\pi$. Insets show the profile structures of the reflected
beam.}

Figures~\rpict{sw_angle_gap}(c,d) show the relative beam shift and
width versus the normalized thickness of the gap medium. The
resonances presented in Figs.~\rpict{sw_angle_gap}(c,d) can be
explained with the help of a simple physics argument. Indeed, when
the gap separating medium 1 and 3 is absent (i.e. $d=0$) or very
small, no surface waves are excited; and the beam shift is
negligible. Increasing the width of the gap medium, we increase
the quality factor of the surface mode, and thus increase the
shift of the reflected beam.  Similarly, for large values of $d$
surface waves are not excited, and the shift of the reflected beam
becomes small again.

To gain a deeper understanding of the physical mechanism for the
large values of the Goos-H\"{a}nchen shift in the case when the
layered structure includes the DNG medium, we calculate the energy
flow distribution and compare it with the results for a
conventional (or right-handed) medium~\cite{Lai:2000-7330:PRE}.
From the analysis of the energy flow structure, we conclude that
the surface wave excited at the interface has a finite extension
and a distinctive vortex-like structure as predicted earlier in
Ref.~\cite{Shadrivov:2003-2713:APL} and other studies. This
surface wave transfers the energy in the negative direction and,
consequently, the energy is reflected from the interface as a
shifted beam.

To make our predictions more realistic, we included the effect of
losses into our analysis, which are always present in DNG
metamaterials. We introduce losses by introducing imaginary parts
into the dielectric permittivity $\epsilon_3$ and magnetic
permeability $\mu_3$. In particular, we take ${\cal I}m(\mu_3)= -
2\cdot 10^{-5}$ and vary the imaginary part of $\epsilon_3$. We
notice that the losses in the DNG medium primarily affect the
surface waves.  Therefore, the major effect produced by the losses
is observed for the strongly shifted beam component.

When the beam is narrow, i.e. its spectral width is large, only a
part of the beam energy is transferred to the surface wave, while
the other part is reflected. This case is shown is
Figs.~\rpict{sw_losses}(a,b). In this case an increase in the
loss, i.e. the increase of the absolute value of ${\cal
I}m(\epsilon_3)$, results in the suppression of the second peak
present in the reflected beam which, as noted above, is due to the
surface wave excitation \cite{Shadrivov:2003-2713:APL}.

Finally, we perform direct numerical FDTD simulations to study the
temporal dynamics of the surface wave excitation.  The FDTD
simulator used in these calculations is described in
~\cite{Ziolkowski:2003-662:OE, Ziolkowski:2003-1596:OE}.
The DNG medium is modelled with a lossy Drude model for both the
permittivity and permeability.  The cell size was set at
$\lambda_0/100$ to minimize any effects of the numerical
dispersion associated with the FDTD method.  We launch the beam
with a waist $\lambda_0$ at the incident angle $21.17^{\circ}$ to
observe the backward wave excitation, i.e., this incident angle
corresponds to the resonant surface wave excitation. The medium
parameters are the same as those used in
Fig.~\rpict{sw_angle_gap}(a,b). The intensity of the electric
field at the final step of the numerical simulations is shown in
Fig.~\rpict{sw_fields}(a). In the top part of this figure we
observe the interference of the incident and reflected beams.
Though it is not easy to discern the double-peak structure of the
reflected beam, one can clearly see the surface wave excited at
the boundary between air and DNG media. The fact that the maximum
of surface wave is shifted in the direction opposite to the
direction of the incident wave indicates that the excited surface
wave is backward. The temporal variations of the amplitudes of the
incident and surface waves are shown in
Fig.~\rpict{sw_backward_history}.

\pict{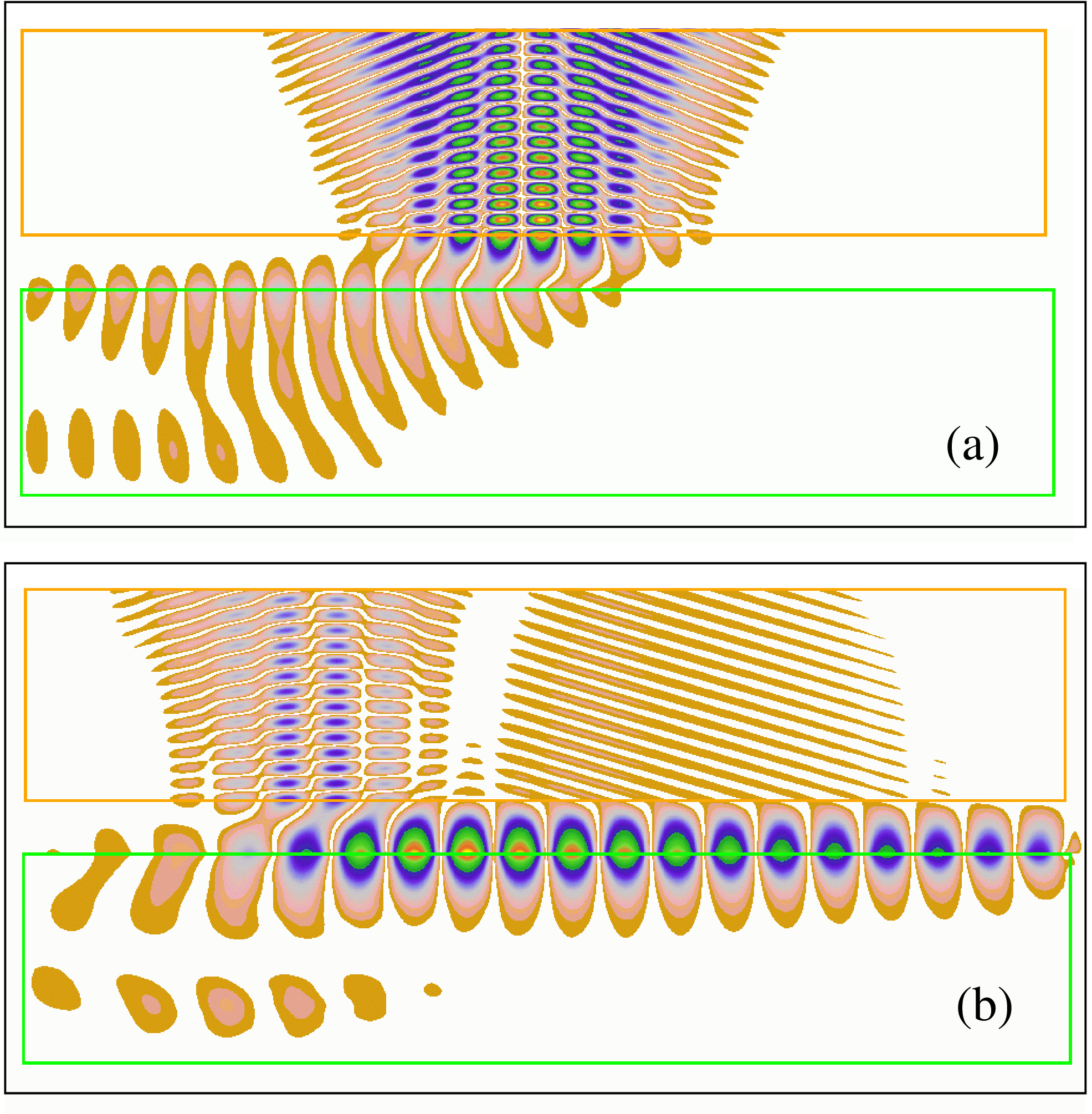}{sw_fields}{Distribution of the electric field
after the excitation of (a) backward surface wave (665K), and (b) forward
surface wave (815K).}
\pict{fig05}{sw_backward_history}{Temporal variation
of the amplitudes of the incident (solid) and surface (dashed)
waves.}

In order to observe the process of the excitation of the forward
surface wave, we take the medium parameters with a different set
of values (see, e.g., Ref.~\cite{Shadrivov:2004-16617:PRE}):
$\epsilon_1 = 12.8$, $\mu_1 = \epsilon_2 = \mu_2 = 1$, $\epsilon_3
= -0.5$, $\mu_3 = - 2$, and $d = \lambda$. Using the theoretical
approach discussed above, we find that for an incident beam having
a waist $\lambda_0$, the resonant excitation of the forward
surface waves should be observed with the incident angle of
$16.32^{\circ}$. The distribution of the electric field intensity
calculated by the FDTD simulator is shown in
Fig.~\rpict{sw_fields}(b). Here, we can identify clearly {\em the
double-beam structure} of the reflected beam discussed above. The
temporal dynamics of the forward wave excitation are similar to
the case of the backward wave. The amplitude of the forward wave
is much higher than the amplitude of the incident wave, in
contrast to the case of the excitation of the backward wave shown
in Fig.~\rpict{sw_backward_history}.

\section{Excitation of slab modes}

Now we consider the five-layer structure geometry shown in
Fig.~\rpict{slab_geom_modes}(a).  The first and fifth slabs have
the material parameters $\epsilon_1$ and $\mu_1$.  There are two
gap slabs with material parameters $\epsilon_2$ and $\mu_2$.  The
middle slab has the material parameters $\epsilon_3$ and $\mu_3$.
Without the slabs one and five (i.e. when $d \rightarrow \infty$)
the structure reduces to an isolated slab.  When this slab is a
DNG medium, it is known to support guided modes. The presence of
the optically dense medium makes these guided slab modes {\em
leaky}~\cite{Tamir:1975-273:NRO}, because these waves can now
tunnel outside the guided region. The dense media on both sides of
the center slab make it possible for such leaky waves to radiate
in both directions.

\pict{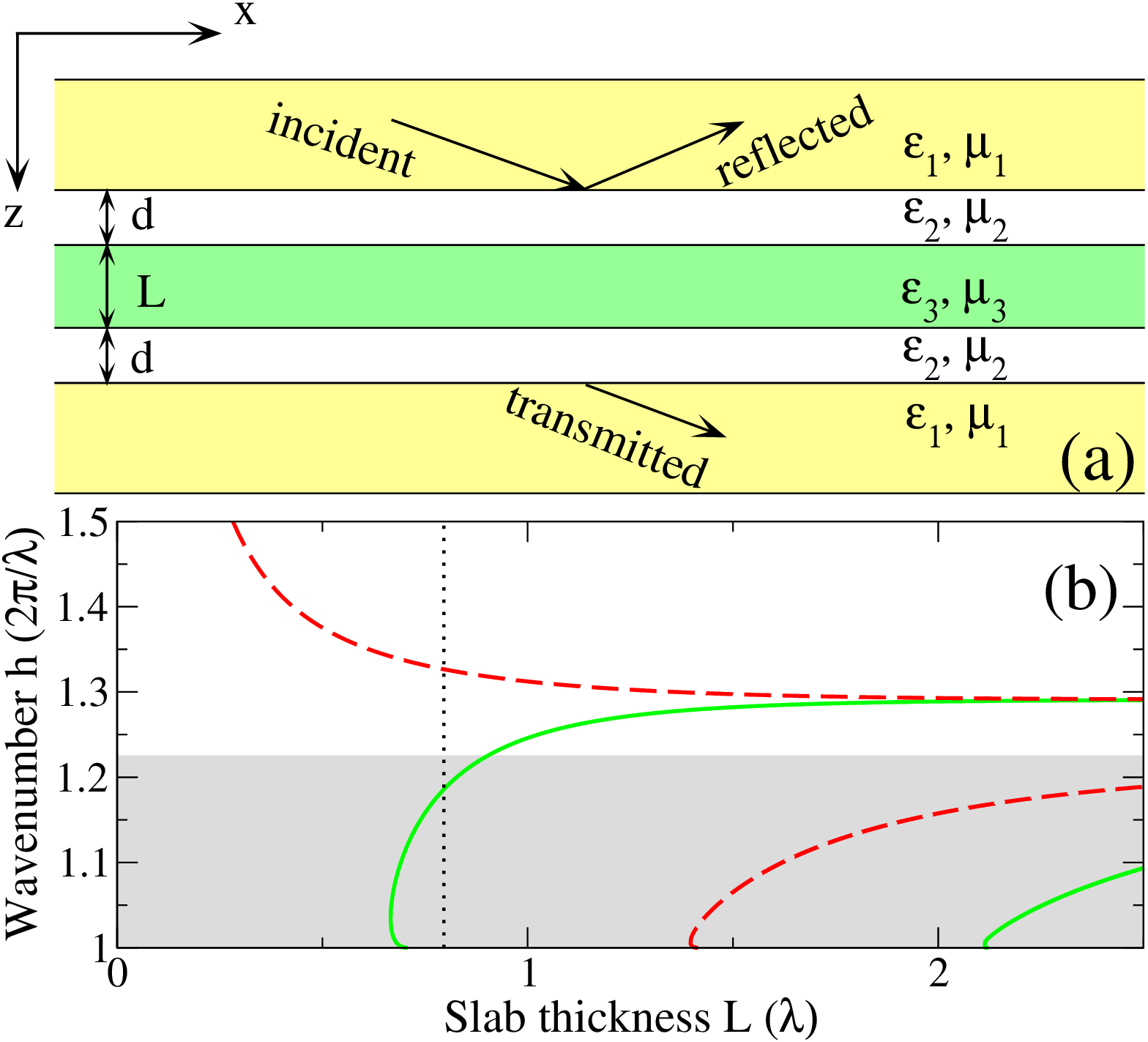}{slab_geom_modes}{(a) Geometry of the
layered structure. (b) Dependence of the normalized wave number
$h$ of the guided modes in the center slab whose thickness is $L$,
for odd (dashed) and even (solid) modes. The vertical dashed line
in the lower figure corresponds to the thickness $L =
5\lambda/2\pi$ used in our calculations.}

For our studies, we use the same parameters as we did for the
three-layered structure discussed in the previous section. The
properties of the guided modes supported by a DNG slab have been
analyzed earlier in Ref.~\cite{Shadrivov:2003-57602:PRE}.
Depending on the thickness of the DNG medium, the slab can support
either one or several guided modes.
Figure~\rpict{slab_geom_modes}(b) shows the dependence of the wave
number of the TE-polarized modes as a function of the slab
thickness.

Reflection and transmission coefficients for the scattering of
monochromatic plane waves by a layered structure can be calculated
with the help of the transfer-matrix method (see, e.g.
Ref.~\cite{Yeh:1988:OpticalWaves}). We take the slab thickness $L
= 5\lambda/2\pi$, so that both symmetric and antisymmetric modes
can exist in this layered structure.  Additionally, we select
angles of incidence so that $k_x$ will be the same as one of the
guided modes, as was discussed above, to achieve large values of
the lateral shift of the reflected beam

\pict{fig07}{slab_angle}{Dependence of the relative shifts of
(a) reflected and (b) transmitted beams versus the angle of
incidence, for $L = 5\lambda_0/2\pi$ and $d = \lambda_0$, and
several values of the waist of the incident beam $a$: $a =
\lambda_0$ (dotted), $a = 5\lambda_0$ (dashed), and $a =
10\lambda_0$ (solid). The vertical lines indicate the position of
the slab eigenmodes. The insert shows an enlargement of the domain
marked by a dashed box in the main figure.}

Figure~\rpict{slab_angle} shows the shift of the reflected and
transmitted beams, calculated with the help of Eq.~\reqt{shift},
as a function of the angle of incidence.  There is a distinctive
resonant behavior for this shift. For a wide beam, the resonance
maxima correspond to the phase matching condition $k_{x0} = h$;
and, measuring the position of such resonances, we can determine
the thickness of the DNG slab with a precision exceeding a
wavelength. However, the two-peak structure of the resonances
disappears for narrower beams because such beams have a wide
angular spectrum and, hence, both modes are excited
simultaneously.  Moreover, the relative shift of the transmitted
beam can be much larger than that of the reflected beam. This
happens because the transmitted wave is composed only of a beam
emitted by the excited leaky wave whereas the reflected beam
consists of two parts: this leaky wave part and the mirror-like
beam reflected from the structure.  In contrast, the positive
resonances in the low wave number gap regions correspond to the
resonant reflections from that gap.  The resulting fields can also
be treated as excitations of leaky waves that are guided by the
air gaps.

\pict{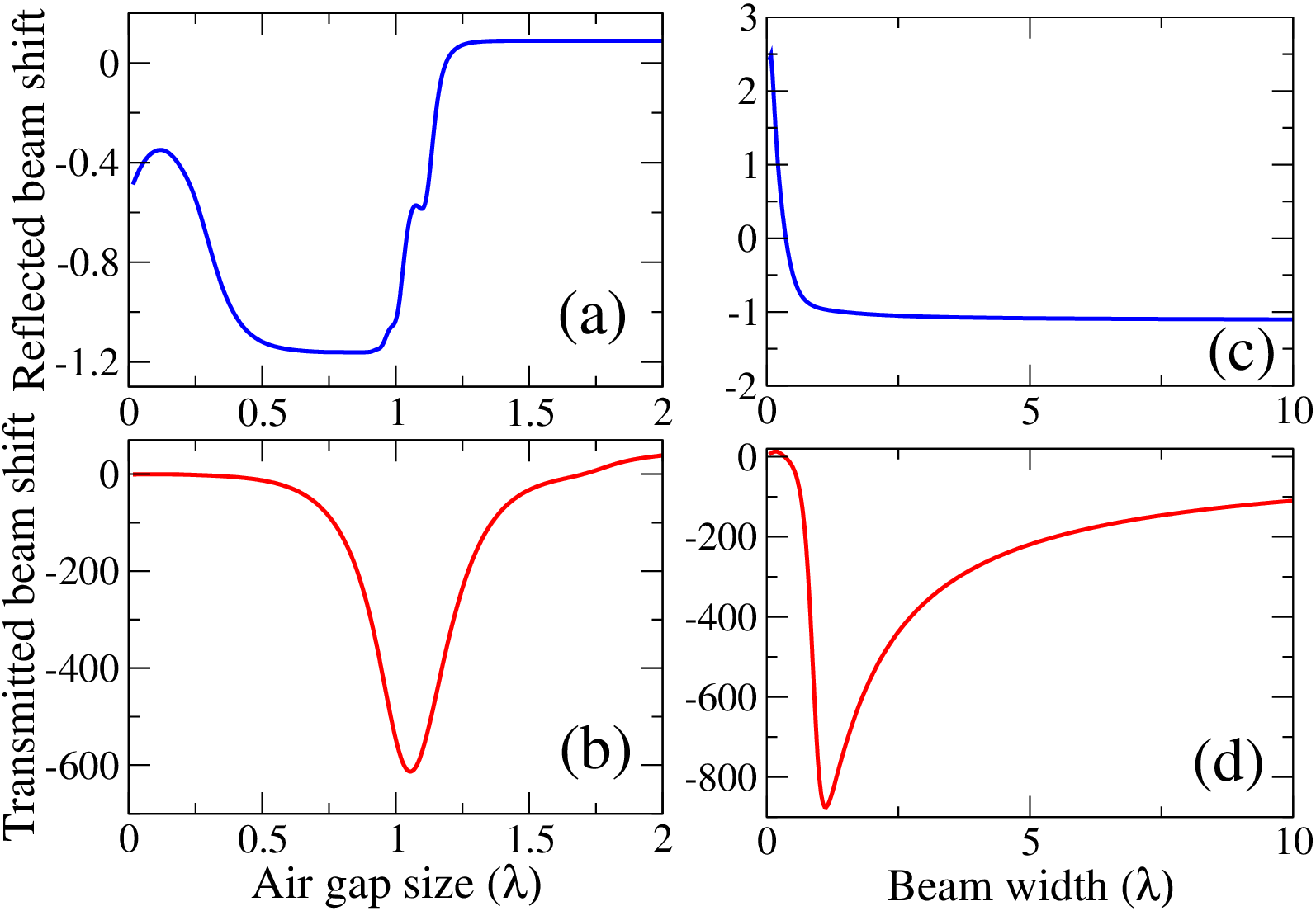}{slab_gap_width}{Dependence of the relative
shift of the (a) reflected and (b) transmitted beams versus the
thickness $d$ of the air gaps between the DNG slab and the
high-index slabs when $L = 5\lambda_0/2\pi$, $a = \lambda_0$, and
$k_{x_0} = 1.186 2\pi/\lambda_0$. Dependence of the relative shift
of the (c) reflected and (d) transmitted beams versus the waist
$a$ of the incident beam when $L = 5\lambda_0/2\pi$, $d =
\lambda_0$, and $k_{x_0} = 1.186 2\pi/\lambda_0$.}

\pict{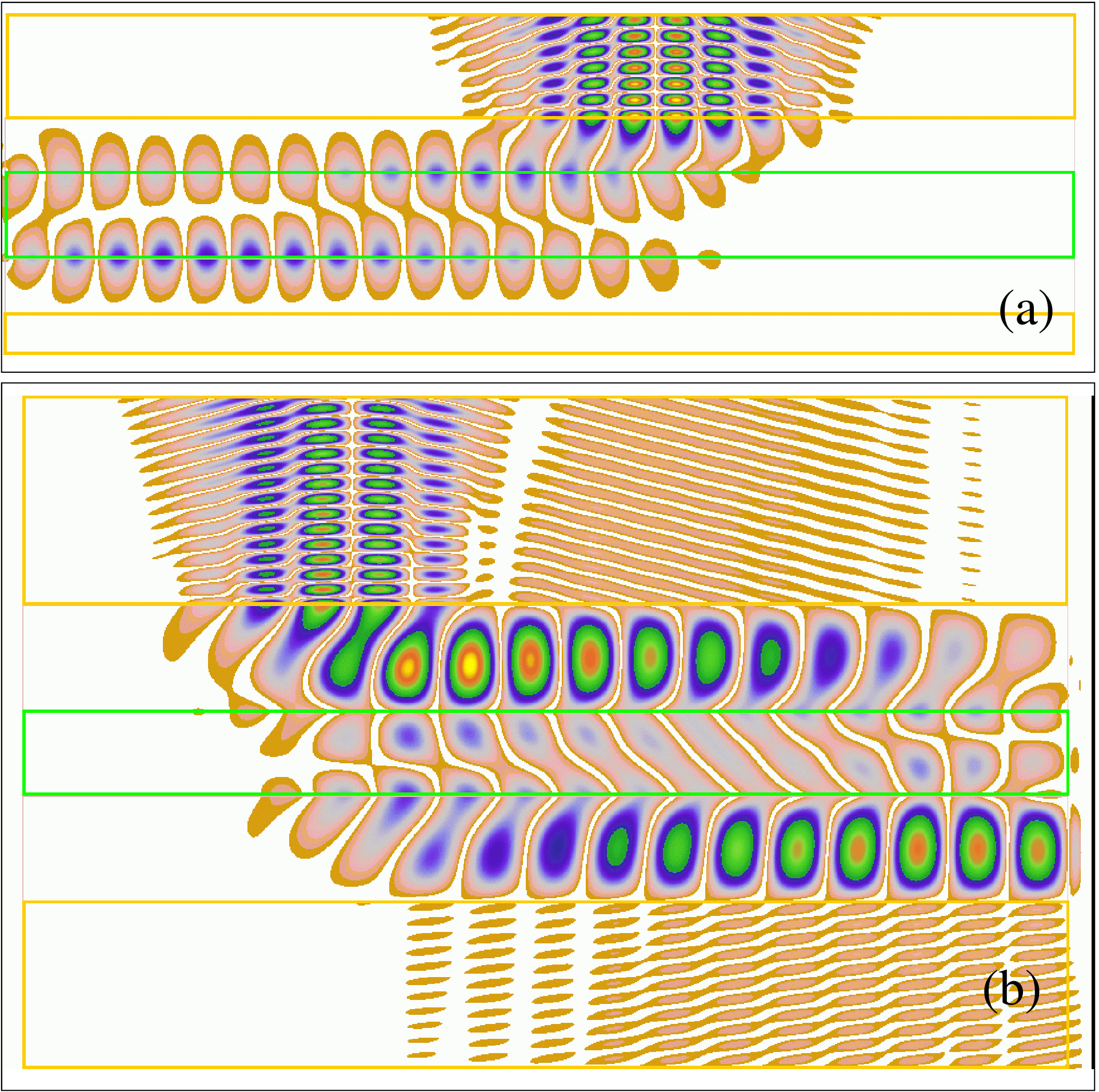}{slab_fields}{Intensity distribution of the
electric field  for the excitation of (a) backward guided waves (430K)
and (b) forward leaky waves guided by the air gaps (1.5M).}
\pict{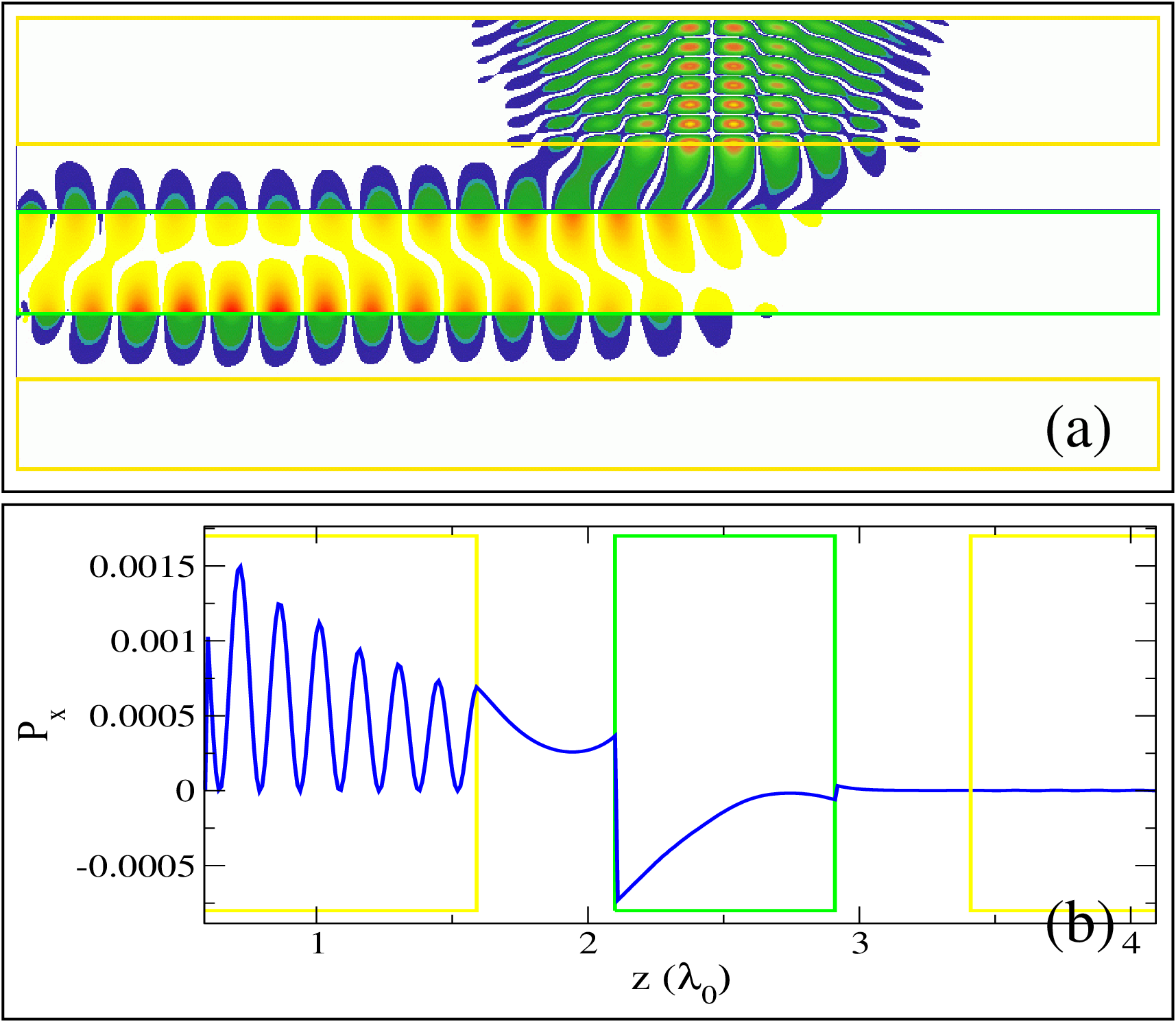}{slab_flows}{(a) (365K) Contour plot of the
$x$-component of the Poynting vector (blue corresponds to positive
values, while yellow corresponds to negative values), (b) Profile
of the $x$-component of the instantaneous Poynting vector as a
function of $z$ (normal to the interfaces) at the middle point of
the simulation domain.}

Figure~\rpict{slab_gap_width} shows the shift of the reflected and
transmitted beams versus the normalized thickness $d$ of the gap
slabs. The resonant-like behavior of these dependencies can be
explained in the same way as was done for the case of surface
waves. If the gap is absent, no leaky waves are excited and,
therefore, the shift of the reflected beam is much smaller then
the incident beam width. If we increase the width of the gap
separating the medium one and three, we increase effectively the
quality factor for the excitation of guided modes, thus increasing
the propagation distance of such waves and, hence, the lateral
shift of the reflected beam. For large widths of the gap, the
reflected beam spectrum is much wider than the spectrum of the
leaky waves and, as a result, only a small part of the beam energy
can be transferred to the guided waves, making the resulting beam
shift small.

Figures~\rpict{slab_gap_width}(c,d) show the dependence of the
beam shift on the width of the incident beam. A change of the
width of the incident beam modifies its spectral extent, thus it
changes the ratio of the energy carried by the mirror-like
reflected and leaky-wave radiated beams.

We have performed direct numerical FDTD simulations of the
temporal dynamics for the excitation of the guided waves in the
five-layer structure.  Samples of these results are shown in
Figs.~\rpict{slab_fields}(a,b) where two snapshots in time of the
electric field intensity distribution at the end of the
simulations are given. In particular, Fig.~\rpict{slab_fields}(a)
shows the excitation of the backward guided waves. The air gap
thickness here was chosen to be $\lambda_0/2$ in order to decrease
the quality factor of the guided modes and to obtain a better
coupling with the radiative modes.  The excited guided wave has a
vortex-like structure of the energy flow, as predicted earlier in
Ref.~\cite{Shadrivov:2003-57602:PRE}. The structure of the
$x$-component of the instantaneous Poynting vector is shown as a
contour plot in Fig.~\rpict{slab_flows}(a). It shows that the
energy inside the DNG slab flows in the direction opposite to
energy flow in dielectrics. Figure~\rpict{slab_flows}(b) presents
the cross-section transverse to the interfaces shown in
Fig.~\rpict{slab_flows}(a) at the middle point of the simulation
domain.  It shows explicitly the negative energy flow inside the
DNG slab waveguide.

Finally, Fig.~\rpict{slab_fields}(b) shows the snapshot in time of
the distribution of the electric field intensity at the end of the
simulation in the case when the excitation of the leaky waves are
guided by the air gaps. The results demonstrate that the electric
field in this case is mostly concentrated in the air gaps. This
explains the positive energy transfer, and the overall positive
shift of the reflected and transmitted beams.

\section{Conclusions}

We have analyzed the scattering of an obliquely incident Gaussian
beam by a multi-layered structure that includes a double-negative
(DNG or left-handed) medium. We have demonstrated that a rich
variety of surface and guided waves supported by these
multi-layered structures having both double-positive (DPS or
right-handed) and DNG media can result in a giant lateral shift of
the beam reflected from it. We have emphasized that this effect is
due to either the resonant excitation of surface waves (surface
polaritons) at the interface between the conventional DPS and the
unconventional DNG materials, or due to the resonant excitation of
guided and leaky modes in the DNG slabs. For the resonant
excitations of guided waves,  the reflected beam has a
well-defined double-peak structure, where one peak represents the
mirror-like reflection, and the second one appears due to a
lateral beam shift from the point of the mirror-like reflection,
and it is produced by the excited surface waves. The lateral beam
shift can be both positive and negative, depending on the type of
the surface waves supported by the structure and excited by the
incoming beam. Many of those predictions hold in the presence of
losses in the DNG material that has been included in our analysis
as well. We have also performed a series of direct
finite-difference-time-domain numerical simulations to model the
temporal dynamics of the beam scattering in both types of the
multi-layered structure guided-wave geometries and have confirmed
the major effects predicted analytically for the time harmonic,
stationary problem.

\section{Acknowledgements}

The authors acknowledge a support of the Australian Research
Council and the Australian National University. Alexander Zharov
thanks the Nonlinear Physics Centre at the Australian National
University for a warm hospitality and research fellowship. The
work by Richard Ziolkowski was supported in part by DARPA under
contract No. MDA972-03-100 and by ONR under contract No.
14-04-1-0320.

\end{sloppy}
\end{document}